\documentstyle[12pt,emlines2]{article}
\newcommand{\NP}[1]{Nucl. \ Phys.}
\newcommand{\PL}[1]{Phys. \ Lett.}
\newcommand{\p}[1]{\partial}
\newcommand{\PRL}[1]{Phys.\ Rev.\ Lett. }
\newcommand{\MPL}[1] { Mod. Phys. Lett. }
\newcommand{\IJMP}[1] { Int. J. Mod. Phys. }
\begin{document}

\title{
Photon splitting cascades  and a new statistics}
\author{
L. Accardi,   I.Ya. Aref'eva
\thanks{
Steklov Mathematical
Institute, Gubkin St.8, GSP-1, 117966, Moscow, Russia}
  and I.V.Volovich
\thanks{
Steklov Mathematical
Institute, Gubkin St.8, GSP-1, 117966, Moscow, Russia}
\\
{\it Centro Vito Volterra Universita di Roma Tor Vergata}}

\maketitle

\begin {abstract}

Photon splitting cascades in a magnetic field are considered. It is shown, in
the stochastic limit approximation, that photons in cascades might form
entangled states (``triphons'') and that they obey not Bose but a new type of
statistics, the so called infinite or quantum Boltzmann statistics.  These
states belong to an interacting Fock space which is a generalization of the
ordinary Fock space. The new photon statistics in principle can be detected
in future   astrophysical experiments such as the planned Integral mission
and also in nonlinear quantum optics.

\end {abstract}
\newpage
\setcounter{equation}{0}

One of the most interesting manifestations of the non--linearity of
Maxwell's equations with radiative corrections is the splitting of a
photon into two in an external magnetic field. In a constant uniform
field, this process occurs with conservation of energy and momentum. The
process was considered by Adler {\it et al.} in the early '70s by using
the Heisenberg--Euler effective Lagrangian \cite{1,2,3}.
Photon splitting was considered as a possible mechanism for the production
of linearly polarized gamma--rays in a pulsar field. Recently the splitting
of photons has found astrophysical applications in the study of annihilation
line suppression in gamma ray pulsars and spectral formation of gamma ray
bursts from neutron stars \cite{7,8}.  Photon splitting cascades have also
been used  in models of soft gamma-ray repeaters, where they soften
the photon spectrum \cite{15,14}. The process of photon splitting is
potentially important in applications to a possible explanation of the origin
of high energy cosmic rays from Active Galactic Nuclei \cite{24}. A
recalculation of the amplitude for photon splitting in a strong magnetic
field has been performed recently in \cite{4,5,6}.

In this note we argue that photon splitting cascades in
the magnetic field might create entangled
states and that photons in cascades obey not Bose but
a new type of statistics, the so called
infinite or quantum Boltzmann statistics. These states are formed from
triples of entangled photons and may be called triphons. They belong to a
generalization of the ordinary Fock space which is called an interacting
Fock space.
Creation and annihilation  operators for infinite (quantum Boltzmann)
statistics satisfy neither Bose nor Fermi commutational relations but
\begin{equation}
                            \label{1}
b({\bf k})b^+({\bf p})=\delta ({\bf k}-{\bf p})
\end{equation}
The relations (\ref{1}) have been considered in several recent
works in quantum field theory.
The second quantized example of infinite statistics has been
discussed by Greenberg \cite{11}
and in the present note it is shown that
this statistics has a physical meaning since it describes photons in
cascades and more generally the dominating diagrams in the long time/week
coupling limit in quantum field thory. The notion of
interacting Fock space was introduced by Accardi and Lu \cite{9} in
nonrelativistic QED and it was related with the role of non--crossing
diagrams in the stochastic limit.  The master field describing
the large $N$ limit in QCD was obtained in \cite{10}, it is quantized
by using the relations (\ref{1});
see \cite{12,13} for a recent discussion of the large $N$ limit.

We will start from a discussion of  the theory of photon splitting
cascades and show the emergence of infinite statistics in this theory
and then discuss its connection with the
stochastic limit of quantum field theory.

In the decay of a photon with momentum ${\bf k}$ into photons with
momentum ${\bf k}_1$ and ${\bf k}_2$, we have conservation of momentum
and energy ${\bf k}={\bf k}_1+{\bf k}_2\ ,$
$\omega({\bf k})=\omega_1({\bf k}_1)+\omega_2({\bf k}_2)$.
For photons in vacuum, in the absence of external fields,
$\omega=\omega_1=\omega_2=k$ and although these two equations have a solution
the decay is forbidden by the invariance under charge
conjugation (Furry's theorem).

In a constant uniform magnetic field ${\bf B}_0$ there are only two
decay processes kinematically allowed, $\gamma_{\|}\to\gamma_\perp+
\gamma_\perp$ and $\gamma_{\|}\to\gamma_{\|}+\gamma_{\perp}$ \cite{2}.
Here the subscripts $\perp$ and $\|$ will denote polarizations of the photon
with respect to the vector ${\bf B}_0$.
More precisely, in  presence of a magnetic field
one has a distictive plane, namely the ${\bf k}{\bf
B}_0$ plane. One takes the linear polarization of the
magnetic field of the photon parallel  and orthogonal  to
this plane as the two independent polarizations of the photon, $\|$ and
$\perp$, respectively.

The vacuum in the presence of the field ${\bf B}_0$ acquires an index
of refraction $n$, and the photon dispersion relation is modified from
$k/\omega=1$ to $k/\omega=n$. The indices of refraction $n_{\|,\perp}$
can be calculated from the Heisenberg--Euler effective lagrangian.
Adler showed that for subcritical fields in the limit
of weak vacuum dispersion only the splitting mode $\|\to\perp+\perp$
operates below pair production threshold.
For weak dispersion $n_\perp=1+{7\over90}\,\beta$ and $n_{\|}=1+{2
\over45}\,\beta$, where $\beta={e^4h\over
m^4c^7}\,B^2_0\sin^2\theta$ and $\theta$ is the angle  between ${\bf
k}$ and ${\bf B}_0$. It is mentioned by Harding {\it et al.} that in
magnetar models of soft gamma repeaters \cite{8}, where supercritical
fields are employed, moderate vacuum dispersion arises. In such a
regime, it is not clear whether Adler's selection rules still endure
since in his analysis higher order contributions to the vacuum
polarization are omitted. In \cite{8} photon cascades are considered for
the case where all three photon splittings modes allowed by CP
invariance are operating. Baier {\it et al.} \cite{5} have found that there
is only one allowed transition $(\|\to\perp+\perp)$ for any magnetic field.
They suggested that a photon cascade could develop only if magnetic
field changes its direction. It seems that the question on the validity
of Adler's rule for a non weak vacuum dispersion deserves a further
study. In this work we consider photon cascades when both kinematically
allowed modes $(\|\to\perp+\perp$ and $\|\to\|+\perp)$ operate.

The interaction operator for the decay $\| \to \perp+\perp$ is known to be
\cite{3}
\begin{equation}
                                  \label{2}
V_1(t)=\lambda_1\int({\bf B}_0{\bf E}_1)({\bf B}_0
{\bf E}_2)({\bf B}_0{\bf B})d^3x,
\end{equation}
where the coupling constant $\lambda_1=13e^6/315\pi^2m^8$ and
magnetic and electric parts of photon field are
\begin{equation}
                     \label{3}
{\bf B}=i(4\pi)^{1/2}{\bf k}\times{\bf e}_{\|}e^{-i({\bf k}{\bf r}-
\omega t)}a_{\|}({\bf k}_1),~~
{\bf E}_1=-i(4\pi)^{1/2}\omega_1{\bf e}_{\perp} e^{i({\bf k}_1
{\bf r}-\omega_1t)}a^+_{\perp}({\bf k})
\end{equation}
and similarly for ${\bf E}_2$.

For the decay $\|\to\|+\perp$ one has a similar
interaction operator with the operator structure
\begin{equation}
                       \label{6}
{\cal A}^{+}(t)=\lambda a^{+}
_{\|}({\bf k}_1)a^+_{\perp}({\bf k}_2)a_{\|}({\bf
k}) e^{-itE}\delta({\bf k}-{\bf k}_1-{\bf k}_2) \end{equation} where
$E=\omega_{\|}({\bf k})-\omega_{\|}({\bf k}_1)-\omega_{\perp}
({\bf k}_2).
$
The coupling constant $\lambda$ in this case can be estimated as
$\lambda
/\lambda_1=\alpha(B_0/B_{cr})^2$, where $\alpha$ is
the fine structure constant, $\alpha=e^2/\hbar c$ and $B_{cr}=m^2
c^3/ e\hbar\,\simeq4.4\times10^{13}$ Gauss.

Let us consider a photon cascade created by a photon with momentum
${{{\bf k}}}$ and polarization $\|$. The photon splits as
$\gamma_{\|}({{{\bf k}}})\to\gamma_{\perp}({\bf k}_1)+\gamma_{\|}
({{{\bf k}}}-{\bf k}_1)$.
Then one has the next generation of splitting:
$\gamma_{\|}({{{\bf k}}}-{\bf k}_1)\to\gamma_{\perp}({\bf k}_2)+
\gamma_{\|}({{{\bf k}}}-{\bf k}_1-{\bf k}_2)$ etc.
After $N$ generations of splitting one gets a cascade with $N$
photons with $\perp $ polarization and momenta
${\bf k}_1,\dots $, ${\bf k}_{N}$ and also one photon with
$\|$ polarization and momentum
${{{\bf k}}}-
{\bf k}_1-\dots-{\bf k}_N$.
An example of a cascade with two generations is shown in Fig 1.
\begin{figure}
\begin{center}
\special{em:
linewidth 0.4pt}
\unitlength 0.30mm
\linethickness{0.4pt}
\begin{picture}(205.00,65.00)
\emline{60.00}{3.00}{1}{143.00}{3.00}{2}
\emline{72.00}{4.00}{3}{69.76}{3.90}{4}
\emline{69.76}{3.90}{5}{68.02}{4.31}{6}
\emline{68.02}{4.31}{7}{66.80}{5.22}{8}
\emline{66.80}{5.22}{9}{66.08}{6.65}{10}
\emline{66.08}{6.65}{11}{66.00}{10.00}{12}
\emline{66.00}{10.00}{13}{66.10}{12.24}{14}
\emline{66.10}{12.24}{15}{65.69}{13.98}{16}
\emline{65.69}{13.98}{17}{64.78}{15.20}{18}
\emline{64.78}{15.20}{19}{63.35}{15.92}{20}
\emline{63.35}{15.92}{21}{60.00}{16.00}{22}
\emline{60.00}{16.00}{23}{57.76}{15.90}{24}
\emline{57.76}{15.90}{25}{56.02}{16.31}{26}
\emline{56.02}{16.31}{27}{54.80}{17.22}{28}
\emline{54.80}{17.22}{29}{54.08}{18.65}{30}
\emline{54.08}{18.65}{31}{54.00}{22.00}{32}
\emline{48.00}{28.00}{33}{45.76}{27.90}{34}
\emline{45.76}{27.90}{35}{44.02}{28.31}{36}
\emline{44.02}{28.31}{37}{42.80}{29.22}{38}
\emline{42.80}{29.22}{39}{42.08}{30.65}{40}
\emline{42.08}{30.65}{41}{42.00}{34.00}{42}
\emline{36.00}{40.00}{43}{33.76}{39.90}{44}
\emline{33.76}{39.90}{45}{32.02}{40.31}{46}
\emline{32.02}{40.31}{47}{30.80}{41.22}{48}
\emline{30.80}{41.22}{49}{30.08}{42.65}{50}
\emline{30.08}{42.65}{51}{30.00}{46.00}{52}
\emline{24.00}{52.00}{53}{21.76}{51.90}{54}
\emline{21.76}{51.90}{55}{20.02}{52.31}{56}
\emline{20.02}{52.31}{57}{18.80}{53.22}{58}
\emline{18.80}{53.22}{59}{18.08}{54.65}{60}
\emline{18.08}{54.65}{61}{18.00}{58.00}{62}
\emline{54.00}{22.00}{63}{54.10}{24.24}{64}
\emline{54.10}{24.24}{65}{53.69}{25.98}{66}
\emline{53.69}{25.98}{67}{52.78}{27.20}{68}
\emline{52.78}{27.20}{69}{51.35}{27.92}{70}
\emline{51.35}{27.92}{71}{48.00}{28.00}{72}
\emline{42.00}{34.00}{73}{42.10}{36.24}{74}
\emline{42.10}{36.24}{75}{41.69}{37.98}{76}
\emline{41.69}{37.98}{77}{40.78}{39.20}{78}
\emline{40.78}{39.20}{79}{39.35}{39.92}{80}
\emline{39.35}{39.92}{81}{36.00}{40.00}{82}
\emline{30.00}{46.00}{83}{30.10}{48.24}{84}
\emline{30.10}{48.24}{85}{29.69}{49.98}{86}
\emline{29.69}{49.98}{87}{28.78}{51.20}{88}
\emline{28.78}{51.20}{89}{27.35}{51.92}{90}
\emline{27.35}{51.92}{91}{24.00}{52.00}{92}
\emline{18.00}{58.00}{93}{18.10}{60.24}{94}
\emline{18.10}{60.24}{95}{17.69}{61.98}{96}
\emline{17.69}{61.98}{97}{16.78}{63.20}{98}
\emline{16.78}{63.20}{99}{15.35}{63.92}{100}
\emline{15.35}{63.92}{101}{12.00}{64.00}{102}
\emline{131.00}{4.00}{103}{128.76}{3.90}{104}
\emline{128.76}{3.90}{105}{127.02}{4.31}{106}
\emline{127.02}{4.31}{107}{125.80}{5.22}{108}
\emline{125.80}{5.22}{109}{125.08}{6.65}{110}
\emline{125.08}{6.65}{111}{125.00}{10.00}{112}
\emline{125.00}{10.00}{113}{125.10}{12.24}{114}
\emline{125.10}{12.24}{115}{124.69}{13.98}{116}
\emline{124.69}{13.98}{117}{123.78}{15.20}{118}
\emline{123.78}{15.20}{119}{122.35}{15.92}{120}
\emline{122.35}{15.92}{121}{119.00}{16.00}{122}
\emline{119.00}{16.00}{123}{116.76}{15.90}{124}
\emline{116.76}{15.90}{125}{115.02}{16.31}{126}
\emline{115.02}{16.31}{127}{113.80}{17.22}{128}
\emline{113.80}{17.22}{129}{113.08}{18.65}{130}
\emline{113.08}{18.65}{131}{113.00}{22.00}{132}
\emline{107.00}{28.00}{133}{104.76}{27.90}{134}
\emline{104.76}{27.90}{135}{103.02}{28.31}{136}
\emline{103.02}{28.31}{137}{101.80}{29.22}{138}
\emline{101.80}{29.22}{139}{101.08}{30.65}{140}
\emline{101.08}{30.65}{141}{101.00}{34.00}{142}
\emline{95.00}{40.00}{143}{92.76}{39.90}{144}
\emline{92.76}{39.90}{145}{91.02}{40.31}{146}
\emline{91.02}{40.31}{147}{89.80}{41.22}{148}
\emline{89.80}{41.22}{149}{89.08}{42.65}{150}
\emline{89.08}{42.65}{151}{89.00}{46.00}{152}
\emline{83.00}{52.00}{153}{80.76}{51.90}{154}
\emline{80.76}{51.90}{155}{79.02}{52.31}{156}
\emline{79.02}{52.31}{157}{77.80}{53.22}{158}
\emline{77.80}{53.22}{159}{77.08}{54.65}{160}
\emline{77.08}{54.65}{161}{77.00}{58.00}{162}
\emline{113.00}{22.00}{163}{113.10}{24.24}{164}
\emline{113.10}{24.24}{165}{112.69}{25.98}{166}
\emline{112.69}{25.98}{167}{111.78}{27.20}{168}
\emline{111.78}{27.20}{169}{110.35}{27.92}{170}
\emline{110.35}{27.92}{171}{107.00}{28.00}{172}
\emline{101.00}{34.00}{173}{101.10}{36.24}{174}
\emline{101.10}{36.24}{175}{100.69}{37.98}{176}
\emline{100.69}{37.98}{177}{99.78}{39.20}{178}
\emline{99.78}{39.20}{179}{98.35}{39.92}{180}
\emline{98.35}{39.92}{181}{95.00}{40.00}{182}
\emline{89.00}{46.00}{183}{89.10}{48.24}{184}
\emline{89.10}{48.24}{185}{88.69}{49.98}{186}
\emline{88.69}{49.98}{187}{87.78}{51.20}{188}
\emline{87.78}{51.20}{189}{86.35}{51.92}{190}
\emline{86.35}{51.92}{191}{83.00}{52.00}{192}
\emline{77.00}{58.00}{193}{77.10}{60.24}{194}
\emline{77.10}{60.24}{195}{76.69}{61.98}{196}
\emline{76.69}{61.98}{197}{75.78}{63.20}{198}
\emline{75.78}{63.20}{199}{74.35}{63.92}{200}
\emline{74.35}{63.92}{201}{71.00}{64.00}{202}
\emline{5.00}{3.00}{203}{66.00}{3.00}{204}
\emline{143.00}{3.00}{205}{193.00}{3.00}{206}
\put(170.00,17.00){\makebox(0,0)[cc]{$\|$}}
\put(205.00,14.00){\makebox(0,0)[cc]{${\bf k}$}}
\put(118.00,38.00){\makebox(0,0)[cc]{${\bf k}_{1}$}}
\put(93.00,63.00){\makebox(0,0)[cc]{$\perp$}}
\put(35.00,63.00){\makebox(0,0)[cc]{$\perp$}}
\put(60.00,38.00){\makebox(0,0)[cc]{${\bf k}_{2}$}}
\put(14.00,17.00){\makebox(0,0)[cc]{$\|$}}
\end{picture}
\end{center}
\label{Fig1}
\caption{Cascade with two generations}
\end{figure}
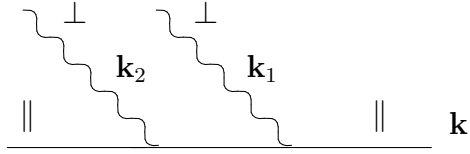
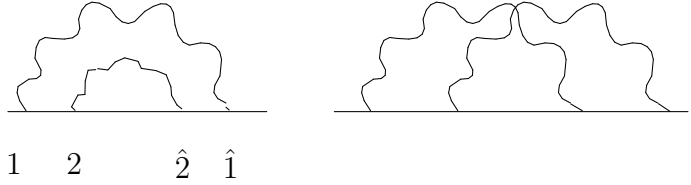
\begin{figure}
\begin{center}
\special{em:
linewidth 0.4pt}
\unitlength 0.60mm
\linethickness{0.4pt}
\begin{picture}(166.00,48.34)
\emline{19.34}{21.67}{1}{17.93}{23.84}{2}
\emline{17.93}{23.84}{3}{17.34}{25.64}{4}
\emline{17.34}{25.64}{5}{17.56}{27.06}{6}
\emline{17.56}{27.06}{7}{20.00}{28.67}{8}
\emline{20.00}{28.67}{9}{21.71}{28.79}{10}
\emline{21.71}{28.79}{11}{22.59}{29.47}{12}
\emline{22.59}{29.47}{13}{22.65}{30.69}{14}
\emline{22.65}{30.69}{15}{21.34}{33.34}{16}
\emline{21.34}{33.34}{17}{21.59}{35.73}{18}
\emline{21.59}{35.73}{19}{22.34}{37.25}{20}
\emline{22.34}{37.25}{21}{23.59}{37.89}{22}
\emline{23.59}{37.89}{23}{25.34}{37.67}{24}
\emline{25.34}{37.67}{25}{27.82}{37.52}{26}
\emline{27.82}{37.52}{27}{29.64}{37.83}{28}
\emline{29.64}{37.83}{29}{30.80}{38.62}{30}
\emline{30.80}{38.62}{31}{31.29}{39.87}{32}
\emline{31.29}{39.87}{33}{31.00}{42.00}{34}
\emline{31.00}{42.00}{35}{31.67}{43.95}{36}
\emline{31.67}{43.95}{37}{32.67}{45.18}{38}
\emline{32.67}{45.18}{39}{33.98}{45.70}{40}
\emline{33.98}{45.70}{41}{35.61}{45.51}{42}
\emline{35.61}{45.51}{43}{38.00}{44.34}{44}
\emline{38.00}{44.34}{45}{39.32}{42.46}{46}
\emline{39.32}{42.46}{47}{40.66}{41.34}{48}
\emline{40.66}{41.34}{49}{42.02}{40.99}{50}
\emline{42.02}{40.99}{51}{43.41}{41.41}{52}
\emline{43.41}{41.41}{53}{45.67}{43.67}{54}
\emline{45.67}{43.67}{55}{47.59}{44.82}{56}
\emline{47.59}{44.82}{57}{49.18}{45.38}{58}
\emline{49.18}{45.38}{59}{50.45}{45.33}{60}
\emline{50.45}{45.33}{61}{51.40}{44.68}{62}
\emline{51.40}{44.68}{63}{52.02}{43.43}{64}
\emline{52.02}{43.43}{65}{52.34}{40.67}{66}
\emline{52.34}{40.67}{67}{53.10}{38.35}{68}
\emline{53.10}{38.35}{69}{54.17}{36.87}{70}
\emline{54.17}{36.87}{71}{55.54}{36.20}{72}
\emline{55.54}{36.20}{73}{58.00}{36.67}{74}
\emline{58.00}{36.67}{75}{59.90}{36.53}{76}
\emline{59.90}{36.53}{77}{61.31}{35.88}{78}
\emline{61.31}{35.88}{79}{62.23}{34.73}{80}
\emline{62.23}{34.73}{81}{62.66}{33.06}{82}
\emline{62.66}{33.06}{83}{62.34}{29.34}{84}
\emline{62.34}{29.34}{85}{61.17}{27.33}{86}
\emline{61.17}{27.33}{87}{61.00}{25.67}{88}
\emline{61.00}{25.67}{89}{61.83}{24.33}{90}
\emline{61.83}{24.33}{91}{63.67}{23.34}{92}
\emline{95.67}{21.67}{93}{94.27}{23.84}{94}
\emline{94.27}{23.84}{95}{93.68}{25.64}{96}
\emline{93.68}{25.64}{97}{93.90}{27.06}{98}
\emline{93.90}{27.06}{99}{96.34}{28.67}{100}
\emline{96.34}{28.67}{101}{98.04}{28.79}{102}
\emline{98.04}{28.79}{103}{98.92}{29.47}{104}
\emline{98.92}{29.47}{105}{98.98}{30.69}{106}
\emline{98.98}{30.69}{107}{97.67}{33.34}{108}
\emline{97.67}{33.34}{109}{97.92}{35.73}{110}
\emline{97.92}{35.73}{111}{98.67}{37.25}{112}
\emline{98.67}{37.25}{113}{99.92}{37.89}{114}
\emline{99.92}{37.89}{115}{101.67}{37.67}{116}
\emline{101.67}{37.67}{117}{104.15}{37.52}{118}
\emline{104.15}{37.52}{119}{105.98}{37.83}{120}
\emline{105.98}{37.83}{121}{107.13}{38.62}{122}
\emline{107.13}{38.62}{123}{107.62}{39.87}{124}
\emline{107.62}{39.87}{125}{107.34}{42.00}{126}
\emline{107.34}{42.00}{127}{108.01}{43.95}{128}
\emline{108.01}{43.95}{129}{109.00}{45.18}{130}
\emline{109.00}{45.18}{131}{110.32}{45.70}{132}
\emline{110.32}{45.70}{133}{111.95}{45.51}{134}
\emline{111.95}{45.51}{135}{114.34}{44.34}{136}
\emline{114.34}{44.34}{137}{115.65}{42.46}{138}
\emline{115.65}{42.46}{139}{116.99}{41.34}{140}
\emline{116.99}{41.34}{141}{118.36}{40.99}{142}
\emline{118.36}{40.99}{143}{119.74}{41.41}{144}
\emline{119.74}{41.41}{145}{122.00}{43.67}{146}
\emline{122.00}{43.67}{147}{123.92}{44.82}{148}
\emline{123.92}{44.82}{149}{125.51}{45.38}{150}
\emline{125.51}{45.38}{151}{126.78}{45.33}{152}
\emline{126.78}{45.33}{153}{127.73}{44.68}{154}
\emline{127.73}{44.68}{155}{128.35}{43.43}{156}
\emline{128.35}{43.43}{157}{128.67}{40.67}{158}
\emline{128.67}{40.67}{159}{129.43}{38.35}{160}
\emline{129.43}{38.35}{161}{130.50}{36.87}{162}
\emline{130.50}{36.87}{163}{131.88}{36.20}{164}
\emline{131.88}{36.20}{165}{134.34}{36.67}{166}
\emline{134.34}{36.67}{167}{136.23}{36.53}{168}
\emline{136.23}{36.53}{169}{137.64}{35.88}{170}
\emline{137.64}{35.88}{171}{138.56}{34.73}{172}
\emline{138.56}{34.73}{173}{138.99}{33.06}{174}
\emline{138.99}{33.06}{175}{138.67}{29.34}{176}
\emline{138.67}{29.34}{177}{137.50}{27.33}{178}
\emline{137.50}{27.33}{179}{137.33}{25.67}{180}
\emline{137.33}{25.67}{181}{138.17}{24.33}{182}
\emline{138.17}{24.33}{183}{140.00}{23.34}{184}
\emline{115.00}{21.67}{185}{113.60}{23.84}{186}
\emline{113.60}{23.84}{187}{113.01}{25.64}{188}
\emline{113.01}{25.64}{189}{113.23}{27.06}{190}
\emline{113.23}{27.06}{191}{115.67}{28.67}{192}
\emline{115.67}{28.67}{193}{117.38}{28.79}{194}
\emline{117.38}{28.79}{195}{118.26}{29.47}{196}
\emline{118.26}{29.47}{197}{118.31}{30.69}{198}
\emline{118.31}{30.69}{199}{117.00}{33.34}{200}
\emline{117.00}{33.34}{201}{117.25}{35.73}{202}
\emline{117.25}{35.73}{203}{118.00}{37.25}{204}
\emline{118.00}{37.25}{205}{119.25}{37.89}{206}
\emline{119.25}{37.89}{207}{121.00}{37.67}{208}
\emline{121.00}{37.67}{209}{123.49}{37.52}{210}
\emline{123.49}{37.52}{211}{125.31}{37.83}{212}
\emline{125.31}{37.83}{213}{126.47}{38.62}{214}
\emline{126.47}{38.62}{215}{126.96}{39.87}{216}
\emline{126.96}{39.87}{217}{126.67}{42.00}{218}
\emline{126.67}{42.00}{219}{127.34}{43.95}{220}
\emline{127.34}{43.95}{221}{128.33}{45.18}{222}
\emline{128.33}{45.18}{223}{129.65}{45.70}{224}
\emline{129.65}{45.70}{225}{131.28}{45.51}{226}
\emline{131.28}{45.51}{227}{133.67}{44.34}{228}
\emline{133.67}{44.34}{229}{134.99}{42.46}{230}
\emline{134.99}{42.46}{231}{136.33}{41.34}{232}
\emline{136.33}{41.34}{233}{137.69}{40.99}{234}
\emline{137.69}{40.99}{235}{139.08}{41.41}{236}
\emline{139.08}{41.41}{237}{141.34}{43.67}{238}
\emline{141.34}{43.67}{239}{143.26}{44.82}{240}
\emline{143.26}{44.82}{241}{144.85}{45.38}{242}
\emline{144.85}{45.38}{243}{146.12}{45.33}{244}
\emline{146.12}{45.33}{245}{147.06}{44.68}{246}
\emline{147.06}{44.68}{247}{147.68}{43.43}{248}
\emline{147.68}{43.43}{249}{148.00}{40.67}{250}
\emline{148.00}{40.67}{251}{148.76}{38.35}{252}
\emline{148.76}{38.35}{253}{149.83}{36.87}{254}
\emline{149.83}{36.87}{255}{151.21}{36.20}{256}
\emline{151.21}{36.20}{257}{153.67}{36.67}{258}
\emline{153.67}{36.67}{259}{155.56}{36.53}{260}
\emline{155.56}{36.53}{261}{156.97}{35.88}{262}
\emline{156.97}{35.88}{263}{157.89}{34.73}{264}
\emline{157.89}{34.73}{265}{158.32}{33.06}{266}
\emline{158.32}{33.06}{267}{158.00}{29.34}{268}
\emline{158.00}{29.34}{269}{156.83}{27.33}{270}
\emline{156.83}{27.33}{271}{156.67}{25.67}{272}
\emline{156.67}{25.67}{273}{157.50}{24.33}{274}
\emline{157.50}{24.33}{275}{159.34}{23.34}{276}
\emline{30.24}{21.75}{277}{29.47}{22.55}{278}
\emline{29.47}{22.55}{279}{30.88}{25.16}{280}
\emline{30.88}{25.16}{281}{32.40}{25.30}{282}
\emline{32.40}{25.30}{283}{32.48}{28.24}{284}
\emline{32.48}{28.24}{285}{33.11}{30.62}{286}
\emline{33.11}{30.62}{287}{34.82}{30.80}{288}
\emline{35.25}{31.01}{289}{37.46}{30.70}{290}
\emline{37.46}{30.70}{291}{39.08}{32.50}{292}
\emline{39.08}{32.50}{293}{41.20}{33.40}{294}
\emline{41.20}{33.40}{295}{44.19}{32.61}{296}
\emline{44.19}{32.61}{297}{44.87}{31.07}{298}
\emline{44.87}{31.07}{299}{48.13}{30.59}{300}
\emline{48.13}{30.59}{301}{50.36}{29.68}{302}
\emline{50.36}{29.68}{303}{51.43}{26.97}{304}
\emline{51.43}{26.97}{305}{51.56}{25.05}{306}
\emline{51.56}{25.05}{307}{52.81}{22.92}{308}
\emline{52.81}{22.92}{309}{53.88}{22.07}{310}
\emline{15.34}{21.45}{311}{72.67}{21.45}{312}
\emline{63.67}{22.45}{313}{64.34}{21.78}{314}
\emline{87.67}{21.45}{315}{166.00}{21.45}{316}
\emline{159.45}{23.24}{317}{161.99}{21.53}{318}
\emline{140.10}{23.18}{319}{142.63}{21.65}{320}
\put(16.67,10.00){\makebox(0,0)[cc]{$1$}}
\put(30.00,10.00){\makebox(0,0)[cc]{$2$}}
\put(54.00,10.00){\makebox(0,0)[cc]{$\hat {2}$}}
\put(64.67,10.00){\makebox(0,0)[cc]{$\hat {1}$}}
\end{picture}
\end{center}
\label{Fig2}
\caption{Crossing and non-crossing diagrams}
\end{figure}

It is important to notice that we consider cascades with real
photons and therefore the diagram in
Fig.1 is not a Feynman one because all the lines (including an
intermediate one) correspond to
real particles on the mass shell and not to virtual states. From the point of
view of the standard quantum field theory all the lines in the diagram
 are "dressed" lines on the mass shell and
moreover the initial photon $\gamma_{\|}(\bf k)$ is prepared in a special way
such that it undergoes the decay in a finite time.  So we cannot
use the standard $S$-matrix approach and the standard Feynman diagram
technique to describe this process.
 The diagram is also not a diagram in the
non--covariant diagram technique \cite {19} because we have the conservation
of energy at every vertex.  The cascade in Fig.1 may be intuitively described
by the following state \begin{equation} \label{8} |\psi( {{{\bf k}}},{\bf
k}_1 ,{\bf k}_2)\rangle= a^+_{\|}({{{\bf k}}}-{\bf k}_1-{\bf k}_2) f({{{\bf
k}}}-{\bf k}_1,{\bf k}_2) a^+_\perp({\bf k}_2)  f({{{\bf k}}},{\bf k}_1)
a^+_\perp({\bf k}_1)|0\rangle
\end{equation}
where momentum conservation
is build in creation and annigilation operators and energy conservation is
accounted for by the factor $f({{{\bf p}}},{\bf k})= f(\omega_{\|}({{{\bf
 p}}})-\omega_\perp({\bf k})-\omega_{\|}({{{\bf p}}}- {\bf k})) $ where
$f(\omega)$ is a function with  support at $\omega=0$.  As we shall see
below this is not the $\delta$--function but roughly speaking its square
root.  Indeed, the transition amplitude between two cascade states
is given by scalar product
\begin{equation}
\label{9}
\langle\psi({{{\bf
k}}}',{\bf k}'_1,{\bf k}'_2)|\psi({{{\bf k}}}, {\bf k}_1,{\bf k}_2)\rangle=
|f({{{\bf k}}},{\bf k}_1))|^2|f({{{\bf k}}}-{\bf k}_1,{\bf
k}_2)|^2\delta({{{\bf k}}}- {\bf k}')\delta({\bf k}_1-{\bf k}'_1)\delta({\bf
k}_2-{\bf k}'_2)
\end{equation}
Notice that in the scalar product (\ref{9})
only the non--crossing diagram, (Fig.2a) contributes. In fact the
contribution from the crossing diagram in Fig.2b vanishes because of
conservation of energy and momentum.  This is a crucial point where
the difference between our diagrams describing real particles
in intermediate states and the Feynman diagrams having virtual particles
in intermediate states is evidentiated. In the Feynman diagram
technique the amplitude
of emission of the two photons is represented by a sum of two diagrams
differing by the order in which the two photons are emitted. Here we have
only one diagram, Fig.1.

Now let us observe that if in (\ref{8}) we replace  operators
$a^+_\perp({\bf k}_1)$ and $a^+_\perp({\bf k}_2)$ by
the quantum
Boltzmann operators $b_\perp({\bf k}_1)$ and $b_\perp({\bf k}_2)$
satisfying the relations (1),
i.e.  $b_\perp({\bf k})b^+_\perp( {\bf p})=\delta({\bf k}-{\bf p})$,
then it will be
automatically guaranteed that only the non-crossing diagrams survive.
Therefore it is natural to describe cascade wave functions in terms of these
operators.  It is well known that standard free photons are bosons.
Therefore to see the quantum Boltzmann statistics we have to prepare
a special state depending on interaction.
A natural method, leading to this result, is
suggested by the stochastic limit technique.
In fact it is natural to expect
that the cascades with physical intermediate states occur at a time scale
slower than the one occurring in the standard $S$- matrix approach to
multiparticle production.  Notice also that the coupling constant in the
interaction term (\ref{6}) are small.  Therefore we are precisely in the
situation in which  one considers  long time--commulative effects of weak
interactions.  The  stochastic limit captures exactly these effects in the
van Hove limit $\lambda\to0$, $t\to\infty$ so that $\lambda^2t\sim\hbox{
constant }=\tau$ (new time scale) which means that we measure time in units
of $1/\lambda^2$ where $\lambda$ measures the strength of the
self--interaction (proportional to the magnitude of the magnetic field in our
case).  It is remarkable that in this limit the triplets of photons
("triphons")   behave like a  single new quantum field obeying a new
statistics.

Now let us consider the question how one can prepare a state
with the new statistics for photons.  If we would deal with the scattering at
infinite time ($S$--matrix) we simply have to consider two Feynman diagrams
to take into account the Bose statistics of photons.  However in the cascade
we deal with evolution in finite time and the states of photons
$\gamma_\perp({\bf k}_1)$ and $\gamma_\perp({\bf k}_2)$ are prepared in a
special way because they are emitted at time $t_1$ and $t_2$, respectively.
Therefore, there is a reason not to add the second diagram.  There exist a
special procedure which is adequated to this situation. This is so-called
stochastic limit technic.  This limiting procedure is widely used in the
consideration of the long time/weak coupling behaviour of quantum dynamical
systems with dissipation, see for example \cite{ALV,AKV}.

In this procedure one deals with states generated by
products of rescaling interaction in different times
in the interaction picture
${1\over\lambda}\,V\left({t\over \lambda^2}\right)$ in
the limit of  $\lambda \to 0$.
This is connected with anisotropic asymptotics,
see \cite{AV}, where one deals with correlators
$<V(t_{1}/\lambda)...V(t_{n}/\lambda)>$
when $\lambda \to 0$.
The equation for the evolution operator in interaction picture reads
$${dU^{(\lambda)}(t)\over dt}\,=-i\lambda V(t)U^{(\lambda)}(t)$$
where $\lambda$ is the coupling constant. In the stochastic
approximation one replaces $U^{(\lambda)}(t)$ to another operator $U(t)$
$$U^{(\lambda)}(t)\approx{\cal U}(\lambda^2t)$$
where ${\cal U}(t)$ is obtained by performing the van Hove rescaling
of time $t\to t/\lambda^2$ and taking the limit $\lambda\to0$. Then for
the limiting evolution operator ${\cal U}(t)=\lim_{\lambda\to0}
U^{(\lambda)}(t/\lambda^2)$ one gets the equation
$${d\,{\cal U}(t)\over dt}\,=-i{\cal V}(t){\cal U}(t)$$
where
$${\cal V}(t)=\lim_{\lambda\to0}{1\over\lambda}\,V\left({t\over
\lambda^2}\right)$$

For the interaction
Hamiltonian (\ref{6}) we consider the asymptotic behaviour of the collective
operator ${\cal A}_{\lambda}(t)
= \frac{1}{\lambda}
{\cal A}(\frac{t}{\lambda ^{2}})$ and its Hermitian conjugate. We obtain
\begin{equation}
                     \label{11c}
 \lim_{\lambda\to 0}{\cal A}^{+}_{\lambda}(t)=b_t^+B^{+}({\bf k}_1,{\bf
 k}_2,{\bf k}) \delta({\bf k}- {\bf k}_1-{\bf k}_2) \end{equation}
 where
 $$B^{+}({\bf k}_1,{\bf k}_2,{\bf k}) =b^+_{\|}({\bf k}_1) b_{\perp}^+({\bf
k}_2) b_{\|}({\bf k})(2\pi)^{1/2}\delta^{1/2}(E) $$
$$E=\omega_{\|}({\bf
k})-\omega_{\|}({\bf k}_1)-\omega_{\perp} ({\bf k}_2).  $$

and the following commutation relations take place
\begin{equation}
\label{12}
b_{t}b^+_{t'}=\delta(t-t')
\end{equation}
\begin{equation}
                                   \label{13}
b_{\perp}({\bf p}) b^+_{\perp}({\bf p}')=\delta({\bf p}-{\bf p}'),~~~
b_{\|}({\bf p}) b^+_{\|}({\bf p}')=\delta({\bf p}-{\bf p}')
\end{equation}
and also the following relation, which explains out notation
$\delta^{1/2}(E) $
\begin{equation}
                                     \label{14}
B({\bf k}_1,{\bf k}_2,{\bf k})B^+({\bf k}'_1,
{\bf k}'_2,{\bf k}')=2\pi\delta(E)\delta({\bf k}_1-
{\bf k}_1)\delta({\bf k}_2-{\bf k}'_2)\delta({\bf k}-{\bf k}')
b^{+}_{\|}(\bf k)b_{\|}(\bf k)
\end{equation}

The relation (\ref{11c}) is understood in the sense of convergence of matrix
elements,
\begin{equation}
                 \label{17}
\langle 0|a_{\|}({{{\bf p}}}_1){\cal A}^\#_\lambda(t_1)\dots{\cal
A}^\#_\lambda(t_n)a^+_{\|}({{{\bf p}}}_2)|0\rangle
\to _{\lambda\to 0}
\langle 0|b_{\|}({{{\bf p}}}_1)B^\#_{t_1}\dots
B^\#_{t_n}b^+_{\|}({{{\bf p}}}_2)|0\rangle
\end{equation}
where ${\cal A}^\#$ means
${\cal A}$ or ${\cal A}^+$

Relations (\ref{13}) define the free or Boltzmann
commutation relations.
Notice the {\it Boltzmannian white noise}
relation (\ref{12}), which makes our model particularly suitable for Monte
Carlo simulations.
The origin for arising these new commutation relations lies in
the fact that the crossing diagrams in the computation of the matrix element
(\ref{17}) are supressed in the weak coupling /large time limit.
The presence of the $\delta (E)$-factor has two important physical
consequences. First,
the commutation relations for the $B^\#$  are not  a consequence of
the corresponding relations for $b^{\#}_{\|}$  and $b^{\#}_{\perp}$:
the three photons are entangled into a single new object
(triphon). Second, the triphon creation and annihilation operators
$B^{\#}$  operate not on the usual Fock space but in
interacting Fock space .

Let us illustrate on the example of four triphons that only diagrams
with non--crossing lines survive in the limit $\lambda\to 0$. An arbitrary
diagram with four "triphons"  schematically
can be written as $${1\over\lambda^4}\,\int \exp
\{\frac{i}{\lambda^2}\sum(\pm)E^{(i)}t_i\} \phi(t,p,k)dk\prod dt_i$$ Here
$E^{(i)}$ are the same as in (\ref{6}) with $k^{(i)}$ being momenta of line
coming in and $k^{(i)}_1, k^{(i)}_2$ momenta of lines coming out from
$i$--vertex; $\pm$ correspond to vertex (\ref{6}) and its
complex conjugated; $p$ are
external momenta and $k$ are two independent momenta;
$\phi $ accomulates all form-factors and test functions.
In general, the sets
of momenta corresponding to different verticies are different. But if there
are two verticies such that momenta coming in the first vertex come out
from the second one and via versa we call these verticies as conjugated
ones (see Fig.2a  where conjugated verticies are denoted by hat). Only
diagrams consistent of pairs of conjugated vertices survive in the limit
$\lambda\to0$. Indeed, making a change of variables $(t_1,t_2,t_{\hat
1},t_{\hat2})\to(\tau_1,\tau_2,t_{\hat1},t_{\hat2})$,
$\tau_1={t_1-t_{\hat1}\over\lambda^2}\,,\ \tau_2={t_2-t_{\hat2}
\over\lambda^2}$
we see that diagram Fig.2a  gives contributions containing the following
factors
$$\delta(E^{(1)}_{\|\to\|+\perp})\delta(E^{(2)}_{\|\to\|+\perp})
\delta(t_1-t_{\hat1})\delta(t_2-t_{\hat2}),$$
where the energy factors $E^{(1)}$, $E^{(2)}$  are not independent due to
momentum conservation, that is typical for interacting Fock space
\cite{9}. For all
others diagrams a similar change of variables do not remove the dependence of
exponent of $\lambda$ that due to fast oscillations produces  zero
contributions.

A photon splits into two not only in a magnetic field
but also in a nonlinear medium. In fact such processes are well known
in nonlinear quantum optics, see for example \cite{WM}.In the
nonlinear process of parametric down conversion a high frequency photon
splits into two photons with frequencies such that their sum equals that of
the high-energy photon.  The two photons  produced in this
process possess quantum correlations and have identical intensity
fluctuations.
The photon pairs generated in parametric down conversion
carry quantum correlations of the Einstein-Podolsky-Rosen type.
Experiments to test Bell inequalities were designed using a
correlated pair of photons.
A two photon cascade was used in
the initial experiment by A.Aspect
to generate the correlated photons but more
recent experiments have used parametric dawn conversion. These experiments
have consistently given predictions of quantum theory and in violation of
"realistic" classical predictions.
In our case the quantum correlations do not come from
superpositions
of spins or polarizations but have a deeper dynamical
origin expressed by the $\delta(E)$-function
in formula (\ref{14}).
It would be very  interesting to extend these experiments to observe
the new statistics considered in this paper.

In conclusion, in this note we have argued that photon cascades in a strong
magnetic field might create entangled states (triphons) which obey not
Bose but the quantum Boltzmann statistics. This prediction is based on the
assumption that both kinematically allowed photon splitting modes operate.
Another assumption is that intermediate photons in a cascade
are physical particles (i.e. on the mass shell) but not
virtual states. This is equivalent to the validity of the stochastic limit
approximation.
In fact both of these assumpions deserve a further study. A better
theoretical understanding of the photon splitting with a non weak dispersion
is required.  From the experimental side new more precise devices such
as the planned Integral misson \cite{7} might significantly advance our
understanding of the fundamental problem of photon statistics.

{\bf Aknowledgement.} I.A. and I.V. are grateful to the Centro Vito Volterra
Universita di Roma Tor Vergata for the kind hospitality.

\end{document}